\documentclass[journal]{IEEEtran}

\ifCLASSINFOpdf
\else
   \usepackage[dvips]{graphicx}
\fi
\usepackage{url}
\usepackage{graphicx}
\usepackage{multirow}
\usepackage{mathtools}
\usepackage{physics}
\usepackage{algpseudocode}
\usepackage{balance}

\usepackage{amsmath, amssymb}
\usepackage[affil-it]{authblk}
\usepackage{enumitem}
\usepackage{booktabs}
\usepackage{tabu}
\usepackage{float}

\usepackage[noadjust]{cite}

\usepackage[usenames,dvipsnames]{xcolor}
\usepackage{footnote}
\usepackage{ctable} 
\usepackage{booktabs}
\usepackage{tabu}
\usepackage[usenames,dvipsnames]{xcolor}
\usepackage{footnote}
\usepackage{ctable} 

\usepackage{algorithm,algorithmicx,algpseudocode}
\usepackage{amsmath}
\usepackage{amsfonts}
\usepackage{amssymb}

\usepackage{makecell}

\begin{document}
\newcommand{\tjnote}{\textcolor{red}}

\title{Auto-Tuning Spectral Clustering for Speaker Diarization Using Normalized Maximum Eigengap}
\author{Tae Jin Park, \IEEEmembership{Member, IEEE}, Kyu J. Han \IEEEmembership{Member, IEEE}, Manoj Kumar and Shrikanth Narayanan, \IEEEmembership{Fellow, IEEE}}
\maketitle
\begin{abstract}
We propose a new spectral clustering framework that can auto-tune the parameters of the clustering algorithm in the context of speaker diarization. The proposed framework uses normalized maximum eigengap (NME) values to estimate the number of clusters as well as the parameters for the threshold of the elements of each row in an affinity matrix during spectral clustering, without any parameter tuning on a development set. Even with this hands-off approach, we achieve comparable or better performance across various evaluation sets than with the traditional clustering methods that use careful parameter tuning and development data. The relative improvement of 17\% in terms of speaker error rate in the well-known CALLHOME evaluation set shows the effectiveness of our proposed auto-tuning spectral clustering. 

\end{abstract}

\begin{IEEEkeywords} Auto-Tuning, Spectral Clustering, Eigengap Heuristic, Speaker Diarization

\end{IEEEkeywords}
\textbf{github.com/tango4j/Auto-Tuning-Spectral-Clustering}

\IEEEpeerreviewmaketitle
\vspace{-1ex}
\section{Introduction}

\IEEEPARstart{S}{peaker} diarization is the problem of identifying ``who spoke when" and assigning speaker identity labels in a given audio stream. In general, speaker diarization systems are comprised of three major parts: Speech segmentation module, speaker embedding extractor, and clustering module. speech segmentation module removes non-speech parts from the audio stream and breaks the speech parts into segments that are supposedly homogeneous in terms of speaker identity. Speaker embedding extractor captures and embeds speaker characteristics in the given segment into a speaker embedding, which is a vector of learned low dimensional representation. Finally, clustering module groups the speaker embeddings from the same speakers into the same clusters.  

Spectral clustering has been widely adopted in numerous speaker diarization studies \cite{ning2006spectral, luque2012, shum2012use, shum2013unsupervised, wang2018speaker, qingjian2019}. Spectral clustering is a graph-based clustering technique that uses an affinity matrix, each element of which is the distance between a pair of speaker embeddings. Throughout the Laplacian matrix computations, the affinity matrix is converted to spectral embeddings, which are clustered by the k-means algorithm \cite{lloyd1982least}. Despite its popularity, spectral clustering has a limitation that its performance is sensitive to the quality of the affinity matrix. Due to the noisy nature of speaker embeddings and distance metrics, it is highly likely for some elements of the affinity matrix to possess noisy signals that could degenerate the clustering process. To address this issue, the spectral clustering algorithms in recent studies employ either a scaling parameter \cite{ning2006spectral, shum2012use, luque2012} or a row-wise thresholding parameter \cite{wang2018speaker} to put different weights across the elements in the affinity matrix. The downside of these approaches is that those parameters for either scaling or thresholding need to be optimized on a development set to obtain the maximum benefit. The burden of such hyper-parameter tuning in spectral clustering would make the generalization of the clustering algorithm harder in unseen testing environments. 

In this paper, we propose a new spectral clustering framework to self-tune the parameters of clustering so that there would be no need for any hyper-parameter tuning with a development dataset. More specifically, our proposed framework estimates both the threshold $p$ for row-wise binarization of a given affinity matrix and the number of clusters $k$. To estimate these parameters without the help of a development set, we use the normalized maximum eigengap (NME) value,  $g_p$, which is dependent on $p$ and can be obtained by the eigengap heuristic \cite{von2007tutorial}. We hypothesize that there exists a piecewise linear relationship between $p$ and $g_p$ and the ratio of $p / g_p$ is a good proxy for diarization error rate (DER). Using this ratio, we can select $p$, where DER would be presumably the lowest. 

To show the experimental evidence, we compare the proposed clustering method with the widely used clustering methods, which need to be optimized on a development set. Our proposed method is compared with the well-known spectral clustering approach \cite{ng2002spectral} appeared in a number of speaker diarization studies \cite{ning2006spectral, shum2012use, luque2012}, and the agglomerative hierarchical clustering (AHC) approach coupled with probabilistic linear discriminant analysis (PLDA) \cite{ioffe2006probabilistic, prince2007probabilistic}, which has also appeared in recent studies \cite{garcia2017speaker, sell2018diarization, snyder_git}. In addition, the performance of the development-set-optimized version of our proposed spectral clustering method is also tested to verify the benefit of our NME-based auto-tuning approach. The experimental results reveal that $p/g_p$ is a good proxy for DER, and the proposed auto-tuning approach can show comparable or even better performance than widely used development-set-optimized clustering algorithms.
\vspace{-1.0ex}
\section{Traditional Spectral Clustering Algorithm}
\subsection{Ng-Jordan-Weiss (NJW) Algorithm}
\label{spectral_clustering_framework}
Spectral clustering is a graph-based clustering technique based on an affinity matrix and its eigenvalues. The affinity matrix is a similarity matrix for a given set of data points, and each element is determined by the distance between a pair of data points in the given input. This algorithm is widely used in diverse fields, such as image segmentation \cite{zelnik2005self}, multi-type relational data \cite{long2006spectral}, and speaker diarization \cite{ning2006spectral, luque2012, shum2012use, shum2013unsupervised, wang2018speaker, qingjian2019}, due to its simple implementation and decent performance. Among many variants of spectral clustering algorithms, the Ng-Jordan-Weiss (NJW) algorithm \cite{ng2002spectral} has been the most widely used for speaker diarization tasks. The NJW algorithm consists of three main steps: creation of the affinity matrix, Laplacian matrix computations, and k-means clustering \cite{lloyd1982least}. To form an affinity matrix, the NJW algorithm employs a kernel method. The similarity measure, which we refer to as $d(\mathbf{w}_i, \mathbf{w}_j)$, between two speaker embeddings from two speech segments is obtained by the cosine similarity measure:
\begin{equation}
\label{eq:cos_sim}
    d(\mathbf{w}_i, \mathbf{w}_j) = \frac{\mathbf{w}_i \cdot \mathbf{w}_j}{\left\lVert \mathbf{w}_i \right\rVert \left\lVert \mathbf{w}_j \right\rVert} .
\end{equation}
Each entry in the affinity matrix $\mathbf{A}$ is defined as follows:
\begin{equation}
a_{ij}=
\begin{cases}
    \text{exp}\big(\frac{(-d(\mathbf{w}_i, \mathbf{w}_j)^{2}}{\sigma^{2}}\big) & \text{if}\; i \neq j \\ 
    0 & \text{if}\; i=j, 
\end{cases}
\end{equation}
where $\sigma$ is a scaling factor that needs to be tuned. $\mathbf{A}$ can be considered as an undirected graph $G=(V,E)$, where $V$ represents vertices and $E$ represents undirected edges. In the NJW algorithm, this affinity matrix $\mathbf{A}$ is normalized with the diagonal matrix $\mathbf{D}$ as follows:
\begin{equation}
    \mathbf{L} = \mathbf{D}^{-\frac{1}{2}} \mathbf{A} \mathbf{D}^{-\frac{1}{2}},
\end{equation}
where $\mathbf{D} = \textrm{diag} \{ d_1, d_2, ..., d_M \}$ and $N$ is the dimension of $\mathbf{A}$. The noramlized matrix $\mathbf{L}$ is used to find the eigenvectors, among which the $k$ largest eigenvectors form a spectral embedding matrix with the size $N \times k$. Each row in this spectral embedding matrix is clustered into one of the $k$ clusters using the k-means clustering. 
\vspace{-1ex}
\subsection{Limitations of the Traditional Spectral Clustering}

Despite its success, the NJW algorithm has inherent limitations in the context of speaker diarization. 
\subsubsection{Sensitivity to the Quality of an Affinity Matrix} 
The similarity values we obtain from distance measures, for example, cosine similarity in (\ref{eq:cos_sim}), for an affinity matrix are merely estimated as well as dependent on how representative the speaker embeddings would be in terms of speaker characteristics. It is likely for some entries in the affinity matrix to have noisy signals that could degenerate the clustering process down the road. Thus, without having a proper scheme to mitigate the effect of such inaccurate information from the affinity matrix, noisy similarity values could lead to a poor clustering result.

\subsubsection{Adaptivity to the Variability in Data} Due to the above issue, there have been a number of schemes proposed in the literature to put different weights across the elements in the affinity matrix. In the previous studies, \cite{wang2018speaker} chose only the entries in each row of the affinity matrix within $p$-percentile, and \cite{ning2006spectral, shum2012use} used scaling factors to control the weights of each element of the affinity matrix. The downside of these approaches is that the parameters for either thresholding or scaling need to be tuned using a development dataset. This could lead to the dependency of the clustering performance upon how to select the development data. Requiring such hyper-parameter tuning would become a burden in generalizing the clustering algorithm in unseen testing conditions. 
\vspace{-1.0ex}
\section{Normalized Maximum Eigengap Analysis} 
In this section, we present the details of our proposed spectral clustering framework using the NME analysis. The procedure is described in Algorithm 1 \footnote{https://github.com/tango4j/Auto-Tuning-Spectral-Clustering}. The following is the itemized description:
\subsection{Steps of the Proposed Clustering Method}
\setlength{\textfloatsep}{5pt}
\begin{algorithm}[t]
\small
\caption{NME-SC algorithm}
\label{alg:loop}
\begin{algorithmic}[1ht]
\Require{Affinity Matrix $\mathbf{A}$} 
\Ensure{Cluster vector $\mathbf{C}$}
\vspace{0.5ex}
\Procedure{\textit{NME-SC}}{$\mathbf{A}$}
    \For{$p \gets 1$ to $P$}
        \State {$\mathbf{A}_p \gets \text{\textit{binarize}}(\mathbf{A}, p)$}
        \State {$\mathbf{\bar{A}}_p \gets (\mathbf{A}_p + \mathbf{A}_{p}^T)/2$}
        \State {$\mathbf{L}_p \gets \text{\textit{Laplacian}}(\mathbf{\bar{A}}_p)$}
        \State {$\mathbf{U}_{p},\mathbf{\Sigma}_{p}, \mathbf{V}_{p}^{T} \gets \text{\textit{SVD}}(\mathbf{L}_p)$}
        \State {$\mathbf{e}_p \gets \text{\textit{eigengap}}(\mathbf{\Sigma}_{p})$}
        \State {$g_{p} \gets \textit{max}(\mathbf{e}_p)/\textit{max}(\mathbf{\Sigma}_p)$}
        \State {$\mathbf{r}[p] \gets p/g_{p}$}
    \EndFor
    \State {$\hat{p} \gets \textit{argmin}(\mathbf{r})$}
    \State {$k \gets \textit{argmax}(\mathbf{e}_{\hat{p}})$}
    \State {$\mathbf{S} \gets \mathbf{U}_{\hat{p}}[1,N;1,k]^{T}$}
    \State {$\mathbf{C} \gets \text{\textit{k-means}}(\mathbf{S}, k)$}
    \State \Return {$\mathbf{C}$}
\EndProcedure
\end{algorithmic}
\end{algorithm}

  \subsubsection{Affinity Matrix} Unlike the NJW algorithm, the affinity matrix \textbf{A} in our proposed framework is formed with raw cosine similarity values in (\ref{eq:cos_sim}) without a kernel or a scaling parameter. From all $N$ speech segments in the given input utterance, we get $N^2$ similarity values as below:
  \begin{equation}
      a_{ij} = d(\mathbf{w}_i, \mathbf{w}_j),
  \end{equation}
  where $i$ and $j$ are indexes of the speech segments.
  \subsubsection{$p$-Neighbor Binarization} The cosine similarity values in the affinity matrix $A$ are binarized to either 0 or 1 to mitigate the effect of unreliable similarity values. This can be done by converting the $p$ largest similarity values in each row to 1 while zeroing out the rest of the values. $p$ is an integer, and is determined by the NME analysis described later.
\begin{equation}
\label{eq:A_p}
    \mathbf{A}_p = \text{\textit{binarize}}(\mathbf{A}, p)
\end{equation}
  \subsubsection{Symmetrization} To transform the affinity matrix $\mathbf{A}_p$ into an undirected adjacency matrix in a graph theory perspective, we perform symmetrization by taking an average of the original and the transposed version of $\mathbf{A}_p$ as follows: 
  \begin{equation}
  \label{eq:A_p_hat}
    \mathbf{\bar{A}}_p =\frac{1}{2}(\mathbf{A}_p + \mathbf{A}_{p}^{T}).
  \end{equation}
  \subsubsection{Laplacian} We use the unnormalized matrix for the Laplacian matrix computations \cite{von2007tutorial} in the following:
\begin{equation}
\begin{split}
    d_{i} &= \sum_{k=1}^{N} a_{ik} \\
    \mathbf{D}_{p} &=\text{diag}\{d_{1}, d_{2}, ..., d_{N}\} \\
    \mathbf{L}_{p} &= \mathbf{D}_{p} - \mathbf{\bar{A}}_{p},
\end{split}
\end{equation}
where $N$ is the size of the matrix $\mathbf{\bar{A}}_p \in \mathbb{R}^{N \times N}$ in (\ref{eq:A_p_hat}).

  \subsubsection{Singular Value Decomposition (SVD)} Perform SVD to obtain eigenvalues for the Laplacian matrix $\mathbf{L}_p$:
     \begin{equation}
     \label{eq:svd}
        \mathbf{L}_{p}=\mathbf{U}_{p}\mathbf{\Sigma}_{p}\mathbf{V}_{p}^{T}.
     \end{equation}
    \subsubsection{Eigengap Vector} Create an eigengap vector $\mathbf{e}_{p}$ as follows, using the eigenvalues from $\mathbf{\Sigma}_{p}$ in (\ref{eq:svd}):
    \begin{equation}
        \label{eq:eig_vec}
            \mathbf{e}_{p} = [\lambda_{p,2}-\lambda_{p,1}, \lambda_{p,3}-\lambda_{p,2}, \cdots, \lambda_{p,N} -\lambda_{p,N-1}],
        \end{equation}
    where $\lambda_{p,i}$ is the $i$-th sorted eigenvalue in ascending order, given $p$ for the binarization process in step \textit{2)}. 
    \vspace{0.3ex}
    \subsubsection{Normalized Maximum Eigengap (NME)}
    The NME analysis is most critical for the auto-tuning part of the proposed spectral clustering algorithm, as we compare the NME values across every $p$ and determine the proper $p$ where DER is presumably minimized. We will discuss this in more detail in Section IV. The NME value $g_p$ for given $p$ is defined as below:
        \begin{equation}
            g_{p} = \frac{\max(\mathbf{e}_p)}{\lambda_{p,N} + \epsilon},
        \end{equation}
    where $\lambda_{p,N} = \max(\mathbf{\Sigma}_{p})$ and $\epsilon$ is a very small value ($\epsilon=1 \times 10^{-10}$). We obtain the ratio $r(p)$ between the pruning threshold $p$ for the row-wise binarization and the NME value $g_p$:
    \begin{equation}
            \label{r_p}
            r(p) = \frac{p}{g_{p}}.
        \end{equation}
    \subsubsection{Estimation of $p$, $\hat{p}$} 
        The value $r_{p}$ is calculated throughout every $p \in \mathbb{N} \cap [1,P]$ and stored in the list $\textbf{r}$ as below:
        \begin{equation}
            \textbf{r} = [ r(1), r(2), \cdots, r(P) ].
        \end{equation}
        Based on our observation, which we will discuss in Section IV as well, $r(p)$ is a very good proxy for DER. Thus, we find the value $\hat{p}$ which makes the minimum value of $\textbf{r}$. Consequently, the parameter $\hat{p}$ attempts to minimize DER:
        \begin{equation}\label{eq:hat_p}
            \hat{p} = \text{argmin}(\mathbf{r}).
        \end{equation}
        With this $\hat{p}$, we estimate the number of clusters $k$:
        \begin{equation}
            k = \text{argmax}(\mathbf{e}_{\hat{p}}).
        \end{equation}
        
    \subsubsection{Spectral Embedding} We take the smallest $k$ eigenvalues and their corresponding eigenvectors to obtain the matrix of $k$-dimensional spectral embeddings $\mathbf{S} \in \mathbb{R}^{k \times N}$:
    
     \begin{equation} 
        \mathbf{S}= \mathbf{U}_{\hat{p}}[1,N;1,k]^{T} = [s_{1}, s_{2},...,s_{N}].
    \end{equation}
    \subsubsection{k-means Clustering} We use the k-means clustering algorithm \cite{lloyd1982least} to obtain $k$ clusters from $\mathbf{S}$.
\vspace{-0ex}
\section{Validation of NME Analysis}
Since our approach of pruning the graph connections of the affinity matrix based on the $p$-neighbor binarization scheme is heavily dependent on the value of $p$, an in-depth analysis is needed for the relationship between the NME value $g_p$ and the pruning parameter $p$.
\vspace{-2ex}
\subsection{Eigengap and the Purity of Clusters}
 It has been known that the size of the eigengap can be used as a quality criterion for spectral clustering \cite{von2007tutorial}. The relation of the size of the eigengap to the purity of clusters has been investigated in \cite{von2007tutorial, stewart1990matrix} using the perturbation theory, more specifically the Davis-Kahan theorem. In this work, we use the NME value $g_p$ to gauge the purity of the clusters, as the purity is directly linked to speaker diarization performance. In doing so, we search the most probable $k$ and the most adequate $p$ altogether using the eigenvalues.
\vspace{-1.0ex}
\begin{figure}[t]
\label{fig:gp_vs_nme}
\vspace{-0.0ex}
\centerline{\includegraphics[width=8.5cm]{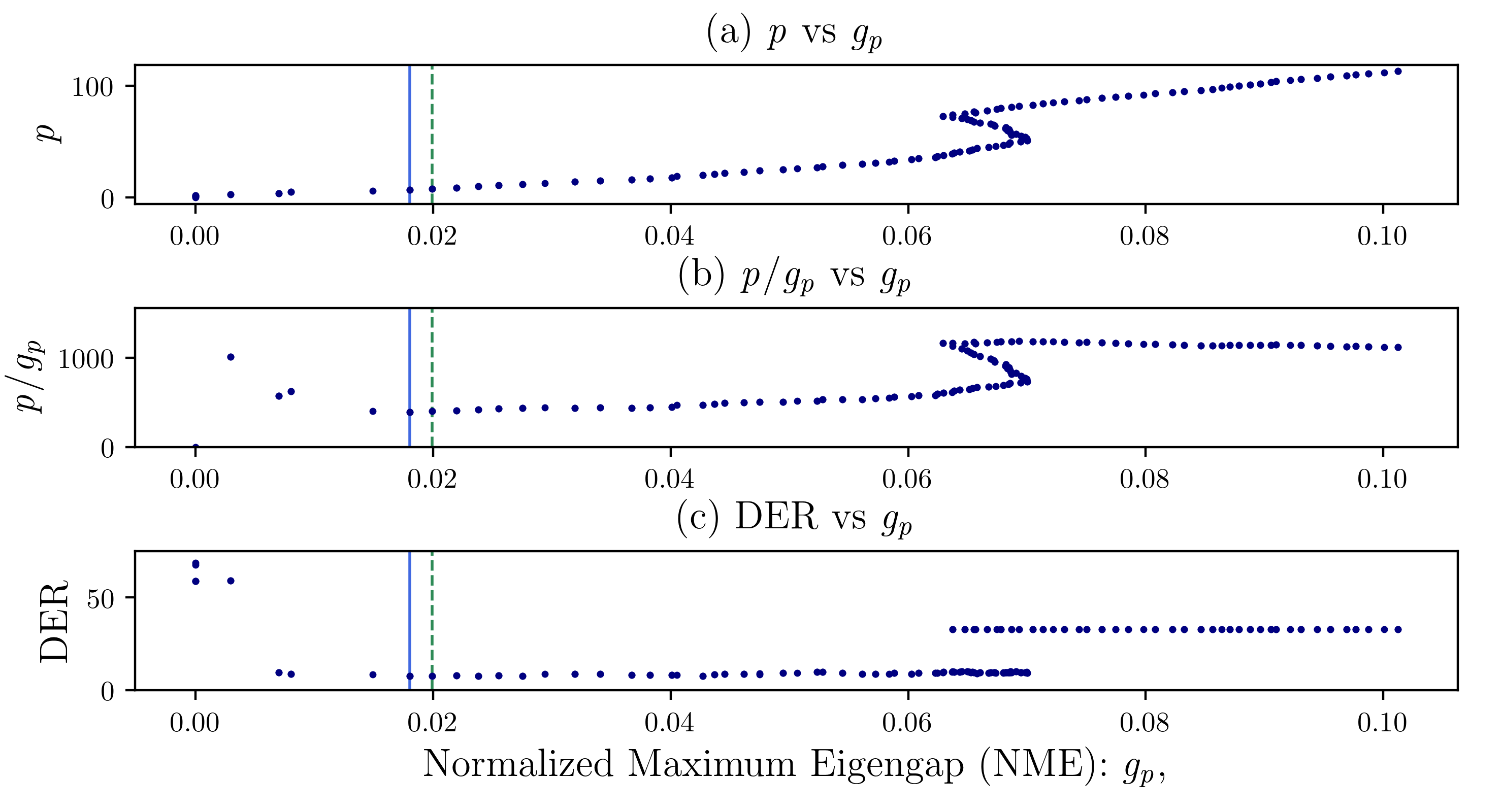}}
\vspace{-1.5ex}
\caption{An example plot that shows the relationship between the NME value $g_p$ and the binarization parameter $p$. The plot of ratio between $p$ and $g_p$ in (b) is reflecting the trend of the DER plot in (c).}
\end{figure}

\subsection{$p$ Versus $g_p$}
Having a higher $p$ value in an affinity matrix $\mathbf{A}$ generally leads to a larger $g_p$ value with the higher purity measure of the clusters, since the graph gets more connections within each cluster. However, since all the connections have the equal weight of 1, an excessive amount of connections (i.e., high $p$ value) gives rise to a poor estimation of the number of clusters followed by a poor diarization result, although it gives a high $g_p$ value. This can be easily understood by thinking of an affinity matrix whose elements are all equal to 1, which would always yield only one cluster regardless of the actual number of clusters. As depicted in Fig.1(a), we see a gradual increase of $g_p$ as $p$ increases while this tendency stops around at $p=50$ in Fig.1(a). As we increase $p$ even more from $p=50$, the estimated number of clusters drops and $g_p$ increases again, meaning that we get a higher $g_p$ value with a smaller estimated number of clusters.   
\begin{table*}[t]
\caption{Experimental results with the Oracle SAD}
\vspace{-3.0ex}
\small
\label{tab:oracle_SAD}
\begin{center}
 \begin{tabular}{ r | c | c | c | c | c }
\Xhline{3\arrayrulewidth}
 & COS+NJW-SC  & COS+AHC & PLDA+AHC & COS+B-SC & COS+NME-SC  \\
 \hline
\textbf{Oracle SAD} & Spk. Err. (DER) & Spk. Err. (DER) & Spk. Err. (DER) & Spk. Err. (DER) &  Spk. Err. (DER) \\
\Xhline{3\arrayrulewidth}
CALLHOME & 24.05 & 21.13 & 8.39 & 8.78 & \textbf{7.29} \\ 
CHAES-eval & 30.31 & 31.99 &24.27 & 4.4  & \textbf{2.48}  \\
CH109 & 13.06 & 29.8 &9.72  & \textbf{2.25} & 2.63 \\
RT03& 6.56 & 5.66 & 1.73  & \textbf{0.88} & 2.21 \\
 \Xhline{3\arrayrulewidth}
\end{tabular}
\end{center}
\vspace{-3.0ex}
\end{table*}

\begin{table*}[t]
\caption{Experimental results with the system SAD}
\vspace{-1.4ex}
\small
\centering
 \begin{tabular}{ r | c c | c c | c c | c c | c c } 
 \Xhline{3\arrayrulewidth}
 \label{tab:system_SAD}
 & \multicolumn{2}{c}{ COS+NJW-SC } & \multicolumn{2}{c}{ COS+AHC } & \multicolumn{2}{c}{ PLDA+AHC } & \multicolumn{2}{c}{COS+B-SC} & \multicolumn{2}{c}{ COS+NME-SC } \\
 \hline
\textbf{System SAD} & DER & Spk. Err. & DER & Spk. Err. & DER & Spk. Err. & DER & Spk. Err. & DER & Spk. Err. \\
\Xhline{3\arrayrulewidth}
 CALLHOME & 26.99 & 20.67 & 20.14 & 13.82 & 12.96 & 6.64 & 13.23 & 6.91 & 11.73 & \textbf{5.41} \\ 
CHAES-eval& 12.04 & 7.73 & 9.96  & 5.85 & 5.52 & 1.45 & 5.07 & 1.00 & 5.04 & \textbf{0.97} \\
CH109 & 5.85 & 1.56 & 28.92 & 24.63 & 6.89 & 2.6 & 5.75 & 1.46 & 5.61 & \textbf{1.32} \\
 
RT03 & 6.42 & 3.88 & 6.24 & 4.7 &   3.53 & 0.99 & 3.1 & \textbf{0.56} & 3.13 & 0.59  \\

\Xhline{3\arrayrulewidth}
\end{tabular}
\vspace{-2.0ex}
\end{table*}

\vspace{-2.0ex}
\subsection{$r(p)=p/g_p$ as a Good Proxy for DER}
Since the excessive amount of connections leads to poor clustering results, $p$ value should be minimized to get an accurate number of clusters, while the $g_p$ value should be maximized to get the higher purity of clusters. Thus, we calculate the ratio $r(p)=p/g_p$ to find the best $p$ value by getting a size of the $p$ value in proportion to $g_p$. It is clearly shown in Fig.1(b) and Fig.1(c) that the ratio of $p$ to $g_p$, $r(p)$ is a very good proxy for DER. As described in (\ref{r_p}),  $r(p)$ indicates the slope in the $p$ versus $g_p$ plot. The lowest $r(p)$ value means that the resulting clusters have the highest purity measure $g_p$ in proportion to $p$. In Fig.1, the solid vertical lines show the estimated point of the lowest DER, while the dotted vertical lines indicate where the actual DER is the lowest.

\vspace{-0.0ex}
\section{Experimental results}

\subsection{Test Setup}
To test the performance of the contribution of the clustering algorithms, we used the same speaker embedding extractor proposed in \cite{snyder2018x, snyder_git} for all the experiments in this work. The evaluation method and metrics followed \cite{fiscus2006rich}. The estimated number of speakers was limited to a maximum of eight speakers for all the experiments. We tested the following five different clustering algorithms: 
\subsubsection{COS+NJW-SC} This is the NJW algorithm in \cite{ng2002spectral} which incorporates the cosine similarity measure. The number of clusters are estimated by the method in \cite{ning2006spectral}.
\subsubsection{COS+AHC} This setup is identical to \cite{snyder2018x, snyder_git}, using the AHC algorithm, except we use cosine similarity instead of PLDA.
\subsubsection{PLDA+AHC} This setup, identical to \cite{snyder2018x, snyder_git}, is AHC coupled with PLDA.  The PLDA model is adapted with each development set.
\subsubsection{COS+B-SC} This is our proposed spectral clustering framework with the $p$-neighbor binarization scheme only, without the NME based auto-tuning approach. i.e., $p$ is optimized on each development set instead of using $\hat{p}$ from (\ref{eq:hat_p}).
\subsubsection{COS+NME-SC} This is our proposed NME-based clustering algorithm, which includes the proposed auto-tuning approach. No hyper-parameter tuning or optimization is done. The $p$ value is searched in the range of $[1, \lfloor\frac{N}{4}\rfloor]$ for each utterance, where $N$ is the number of total segments in a given input utterance.

\vspace{-2.0ex}
\subsection{Datasets}

\subsubsection{NIST SRE 2000 (LDC2001S97)} This dataset is the most widely used for speaker diarization in recent literature, which is referred to as CALLHOME. CALLHOME contains two to seven speakers for each utterance. For the CALLHOME dataset, twofold cross validation is conducted to match the test conditions with \cite{snyder2018x, snyder_git} for all the experiments.
\subsubsection{CALLHOME American English Speech (CHAES) (LDC97S42)}  This is a corpus that only contains English speech data with two to four speakers per each utterance. CHAES is divided into a train, dev, and eval set, and we report the results on the eval set. Both the train set and the dev set are used for parameter turning. The subset of CHAES that only contains two speakers is referred to as CH109 in the literature, and the CH109 dataset is tested by providing the number of speakers ahead to all the tested systems  (i.e., no estimation of the number of speakers involved in CH109). The rest of the utterances in CHAES are used as a dev set for CH109.
\subsubsection{RT03 (LDC2007S10)} RT03 is an English dataset and contains utterances with two to four speakers. We use the split provided by the authors in \cite{wang2018speaker} using only Switchboard utterances. 

\vspace{-2.0ex}
\subsection{Experiments}
\subsubsection{Oracle SAD} Table \ref{tab:oracle_SAD} shows the experimental results with the oracle SAD.  Note that, except for RT03 dataset, NME-SC shows very competitive performances with no parameter tuning at all. The DER of NME-SC is impressive, especially for the CALLHOME dataset, where each utterance has the varying number of speakers, and our proposed auto-tuning approach gains many advantages. 
\subsubsection{System SAD} Table \ref{tab:system_SAD} shows the experimental results for the system SAD. We used the ASpIRE SAD model \cite{povey2011kaldi} that is publicly available. With the system SAD setting, which is closer to scenarios in the wild, NME-SC outperforms all the other methods except for RT03, where NME-SC shows very close performance to dev-set-optimized COS+B-SC method. 

\vspace{-1ex}
\subsection{Discussions}
The performance gain from NJW-SC to B-SC indicates that the $p$-neighbor binarization scheme with the unnormalized Laplacian approach can be effective, as it shows very distinctive performance difference. More importantly, the performance gain from B-SC to NME-SC shows that the value of $p$ can be effectively auto-tuned even without optimizing on a development set. We also see the performance improvement of NME-SC over PLDA+AHC, hinting that our proposed clustering scheme can still get a competitive speaker diarization result without employing PLDA as a distance measure. These all validate the effectiveness of the proposed auto-tuning spectral clustering framework with the NME analysis. 

\vspace{-1ex}
\section{Conclusions}
In this paper, a new framework of auto-tuning spectral clustering was introduced. The experimental results show that our proposed NME-based spectral clustering method is competitive in performance while not requiring any hyper-parameter tuning. It is promising that the proposed method outperformed the widely used AHC method with PLDA. Further work will include a way to theoretically analyze the reason that the ratio of the tuning parameter $p$ to the NME value $g_p$ is a proxy for DER and how it could be generalized in various data on real production systems. 

\bibliographystyle{IEEEtran}
\bibliography{mybib}

\begin{thebibliography}{10}
\providecommand{\url}[1]{#1}
\csname url@samestyle\endcsname
\providecommand{\newblock}{\relax}
\providecommand{\bibinfo}[2]{#2}
\providecommand{\BIBentrySTDinterwordspacing}{\spaceskip=0pt\relax}
\providecommand{\BIBentryALTinterwordstretchfactor}{4}
\providecommand{\BIBentryALTinterwordspacing}{\spaceskip=\fontdimen2\font plus
\BIBentryALTinterwordstretchfactor\fontdimen3\font minus
  \fontdimen4\font\relax}
\providecommand{\BIBforeignlanguage}[2]{{%
\expandafter\ifx\csname l@#1\endcsname\relax
\typeout{** WARNING: IEEEtran.bst: No hyphenation pattern has been}%
\typeout{** loaded for the language `#1'. Using the pattern for}%
\typeout{** the default language instead.}%
\else
\language=\csname l@#1\endcsname
\fi
#2}}
\providecommand{\BIBdecl}{\relax}
\BIBdecl

\bibitem{ning2006spectral}
H.~Ning, M.~Liu, H.~Tang, and T.~S. Huang, ``A spectral clustering approach to
  speaker diarization,'' in \emph{Proc. 9th Int. Conf. Spoken Lang. Process.},
  Sep. 2006, pp. 2178--2181.

\bibitem{luque2012}
J.~Luque and J.~Hernando, ``On the use of agglomerative and spectral clustering
  in speaker diarization of meetings,'' in \emph{Proc. Odyssey: The Speaker and
  Lang. Recognit. Workshop}, Jun. 2012, pp. 130--137.

\bibitem{shum2012use}
S.~Shum, N.~Dehak, and J.~Glass, ``On the use of spectral and iterative methods
  for speaker diarization,'' in \emph{Proc. 13th Annual Conf. Int. Speech
  Commun. Association}, Sep. 2012, pp. 482--485.

\bibitem{shum2013unsupervised}
S.~H. Shum, N.~Dehak, R.~Dehak, and J.~R. Glass, ``Unsupervised methods for
  speaker diarization: An integrated and iterative approach,'' \emph{IEEE
  Trans. Audio, Speech, Lang. Process.}, vol.~21, no.~10, pp. 2015--2028, May
  2013.

\bibitem{wang2018speaker}
Q.~Wang, C.~Downey, L.~Wan, P.~A. Mansfield, and I.~L. Moreno, ``Speaker
  diarization with {LSTM},'' in \emph{Proc. IEEE Int. Conf. Acoust., Speech,
  Signal Process.}, Apr. 2018, pp. 5239--5243.

\bibitem{qingjian2019}
Q.~Lin, R.~Yin, M.~Li, H.~Bredin, and C.~Barras, ``{LSTM} based similarity
  measurement with spectral clustering for speaker diarization,'' in
  \emph{Proc. INTERSPEECH}, Sep. 2019, pp. 366--370.

\bibitem{lloyd1982least}
S.~Lloyd, ``Least squares quantization in {PCM},'' \emph{IEEE Trans. Inf.
  Theory}, vol.~28, no.~2, pp. 129--137, 1982.

\bibitem{ng2002spectral}
A.~Y. Ng, M.~I. Jordan, and Y.~Weiss, ``On spectral clustering: Analysis and an
  algorithm,'' in \emph{Proc. Adv. Neural Inf. Process. Syst.}, Dec. 2002, pp.
  849--856.

\bibitem{ioffe2006probabilistic}
S.~Ioffe, ``Probabilistic linear discriminant analysis,'' in \emph{Proc. Eur.
  Conf. Comput. Vision}, May 2006, pp. 531--542.

\bibitem{prince2007probabilistic}
S.~J. Prince and J.~H. Elder, ``Probabilistic linear discriminant analysis for
  inferences about identity,'' in \emph{Proc. IEEE 11th Int. Conf. Comput.
  Vision}, Oct. 2007, pp. 1--8.

\bibitem{garcia2017speaker}
D.~Garcia-Romero, D.~Snyder, G.~Sell, D.~Povey, and A.~McCree, ``Speaker
  diarization using deep neural network embeddings,'' in \emph{Proc. IEEE Int.
  Conf. Acoust., Speech, Signal Process.}, Mar. 2017, pp. 4930--4934.

\bibitem{sell2018diarization}
G.~Sell, D.~Snyder, A.~McCree, D.~Garcia-Romero, J.~Villalba, M.~Maciejewski,
  V.~Manohar, N.~Dehak, D.~Povey, S.~Watanabe \emph{et~al.}, ``Diarization is
  hard: Some experiences and lessons learned for the {JHU} team in the
  inaugural {DIHARD} challenge.'' in \emph{Proc. INTERSPEECH}, Sep. 2018, pp.
  2808--2812.

\bibitem{snyder_git}
D.~Snyder, ``Callhome diarization recipe using x-vectors,'' Github, May 4,
  2018. [Online]. Available:
  \url{https://david-ryan-snyder.github.io/2018/05/04/model_callhome_diarization_v2.html},
  [Accessed Oct. 9, 2019].

\bibitem{zelnik2005self}
L.~Zelnik-Manor and P.~Perona, ``Self-tuning spectral clustering,'' in
  \emph{Proc. Adv. Neural Inf. Process. Syst.}, Dec. 2005, pp. 1601--1608.

\bibitem{long2006spectral}
B.~Long, Z.~M. Zhang, X.~Wu, and P.~S. Yu, ``Spectral clustering for multi-type
  relational data,'' in \emph{Proc. 23rd Int. Conf. Mach. Learn.}, Jun. 2006,
  pp. 585--592.

\bibitem{von2007tutorial}
U.~Von~Luxburg, ``A tutorial on spectral clustering,'' \emph{Statist. and
  Comput.}, vol.~17, no.~4, pp. 395--416, 2007.

\bibitem{stewart1990matrix}
G.~W. Stewart and J.~Sun, \emph{Matrix Perturbation Theory}.\hskip 1em plus
  0.5em minus 0.4em\relax Boston, MA, USA: Academic Press, 1990.

\bibitem{snyder2018x}
D.~Snyder, D.~Garcia-Romero, G.~Sell, D.~Povey, and S.~Khudanpur, ``X-vectors:
  Robust {DNN} embeddings for speaker recognition,'' in \emph{Proc. IEEE Int.
  Conf. Acoust., Speech, Signal Process.}, Apr. 2018, pp. 5329--5333.

\bibitem{fiscus2006rich}
J.~G. Fiscus, J.~Ajot, M.~Michel, and J.~S. Garofolo, ``The rich transcription
  2006 spring meeting recognition evaluation,'' in \emph{Proc. Int. Workshop
  Mach. Learn. Multimodal Interaction}, May 2006, pp. 309--322.

\bibitem{povey2011kaldi}
D.~Povey, A.~Ghoshal, G.~Boulianne, L.~Burget, O.~Glembek, N.~Goel,
  M.~Hannemann, P.~Motlicek, Y.~Qian, P.~Schwarz \emph{et~al.}, ``The {Kaldi}
  speech recognition toolkit,'' in \emph{Proc. IEEE Workshop Autom. speech
  Recognit. and Underst.}, Dec. 2011.

\end{thebibliography}
\balance

\end{document}